\documentclass{aa} 
\usepackage{psfig}

\begin{document}
\def\ntilde{\hbox{\rm n}}
\def\vv{\hbox{\bf v}}
\def\gvec{\hbox{\bf g }}
\def\rvec{\hbox{\bf r }}
\def\svec{\hbox{\bf s }}
\def\vvec{\hbox{\bf v }}
\def\solphys{\hbox{Sol. Phys.}}
\def\aaps{\hbox{A\&AS}}
\def\aap{\hbox{A\&A}}
\def\apj{\hbox{ApJ}}
\def\apjl{\hbox{ApJL}}
\def\nat{\hbox{Nature}}

\title{Modeling an X-ray Flare on Proxima Centauri: 
evidence of two flaring loop components and of two heating
mechanisms at work\thanks{Based on observations obtained with XMM-Newton, an
ESA science mission with instruments and contributions directly funded by ESA
Member states and the USA (NASA)}}
\author{F. Reale\inst{1} \and M. G\"udel\inst{2} \and G. Peres \inst{1} \and
M. Audard\inst{3}}
\institute{Dipartimento di Scienze Fisiche \& Astronomiche, Sezione di
Astronomia, Universit\`a di Palermo, Piazza del Parlamento 1, 
I-90134 Palermo, Italy; \email{reale@astropa.unipa.it},
\email{peres@astropa.unipa.it} \and
Paul Scherrer Institut, W\"{u}renlingen \& Villigen, CH-5232
Villigen PSI, Switzerland; \email{guedel@astro.phys.ethz.ch} \and
Columbia Astrophysics Laboratory, Columbia University, 550 West
120th Street, New York, NY 10027, USA;
\email{audard@astro.columbia.edu}}

\offprints{F. Reale}
\date{Received /Accepted}

\abstract{ We model in detail a flare observed on Proxima Centauri with
the EPIC-PN on board XMM-Newton at high statistics and high time
resolution and coverage. Time-dependent hydrodynamic loop modeling is
used to describe the rise and peak of the light curve, and a large
fraction of the decay, including its change of slope and a secondary
maximum, over a duration of more than 2 hours.  The light curve, the
emission measure and the temperature derived from the data allow us to
constrain the loop morphology and the heating function and to show that
this flare can be described with two components: a major one triggered
by an intense heat pulse injected in a single flaring loop with
half-length $\approx 1.0 \times 10^{10}$ cm, the other one by less
intense heat pulses released after about 1/2 hour since the first one
in related loop systems, probably arcades, with the same half-length.
The heat functions of the two loop systems appear be very similar: an
intense pulse located at the loop footpoints followed by a low gradual
decay distributed in corona. The latter result and the similarity to at
least one solar event (the Bastille Day flare in 2000) indicate that
this pattern may be common to solar and stellar flares, in wide
generality.
}

\titlerunning{Modeling an XMM-Newton Flare on Prox Cen}

\maketitle

\section{Introduction}
\label{sec:intro}

Coronal flares are known to be very complex phenomena, and to involve
multiple coronal structures, multiple spectral bands and multiple
physical mechanisms at a time. Furthermore,
it is very difficult to define a typical coronal flare
pattern (e.g. Golub \& Pasachoff 1997). A ``standard'' classification,
of solar coronal flares divides them into two main categories, based
fundamentally on the topology of the involved structures: compact flares
and long-enduring events (Pallavicini et al.  1977). Compact flares occur
mostly inside single loops whose shape and volume do not change
significantly during the flare. Long-enduring events, instead, occur in
loop arcades, and higher and higher loops are typically involved as the flare
progresses. The arcade footpoints, best seen in the H$\alpha$ line, appear
as two ribbons getting more distant with time. Long-enduring events
are generally more gradual and longer-lasting than compact
flares, but exceptions exist. 

The soft X-ray light curve of flares consists, generally, of a steep
rising phase, a well-defined peak and a slower -- generally exponential
-- decay. A gradual rise or a decay composed by segments with different
e-folding times (e.g.  Osten \& Brown 1999) can also occur.

Stellar flares are spatially unresolved and we have no direct
information on the morphology of the coronal structures involved,
except in the presence of eclipses during the flare (Schmitt \& Favata
1999). The similarity of solar and stellar X-ray flares, however,
suggests that also stellar flares involve plasma confined in closed
structures.

Empirical methods have been developed to infer the size of the flaring
structures from the e-folding decay time of light curves (Kopp \&
Poletto 1984, White et al. 1986, Poletto et al. 1988, van den Oord \&
Mewe 1989, Pallavicini et al. 1990, Hawley et al. 1995, Reale et al.
1997, Reale \& Micela 1998, see Reale 2002 for an extensive review of
these methods).  In the hypothesis of flares occurring inside closed
coronal structures, the decay time of the X-ray emission roughly scales
as the plasma cooling time. In turn, the cooling time scales with the
length of the structure which confines the plasma: the longer the
decay, the larger is the structure (e.g.  Haisch 1983). A loop
thermodynamic decay time has been derived (van den Oord \& Mewe 1989,
Serio et al.  1991) as:

\begin{equation}
\tau_{th}=\frac{120 L_9}{\sqrt{T_7}}
\label{eq:tauserio}
\end{equation}

\noindent
where $L_9$ and $T_7$ are the loop half-length and the maximum
temperature of the flaring plasma, in units of $10^9$ cm and $10^7$ K,
respectively. The timescale above is derived under the hypothesis of 
impulsive heat released at the beginning of the flare. However, a
significant heat released during the decay may increase the decay time,
and therefore lead to overestimate the loop length, if
not correctly diagnosed (Reale et al. 1997, Reale 2002).  By means of
extensive hydrodynamic simulations of decaying flaring loops, Reale et
al. (1997) derived an empirical formula for the loop length, combining
the information from the light curve and the trajectory of the flare in
the density-temperature diagram \footnote{The square root of the
emission measure can been used as proxy of the density to construct the
density-temperature diagram.}:

\begin{equation}
L_9 = \frac{\tau_{LC}\sqrt{T_7}}{120f(\zeta)} ~~~~~~~~ f(\zeta) \geq 1 
\label{eq:lreale}
\end{equation}

\noindent
where $\tau_{LC}$ is the decay time derived from the light curve. This
formula can be obtained from the expression of the loop thermodynamic
cooling time (Eq.\ref{eq:tauserio}), but includes a non-dimensional
correction factor $f(\zeta)$, larger than one (i.e. a shorter loop
length) if significant heating is present during the decay.  The slope
$\zeta$ of the decay path in the density-temperature diagram (Sylwester
et al. 1993) is maximum ($\sim 2$) if the heating is negligible in the
decay and minimum ($\sim 0.5$) -- the slope of the loci of the
hydrostatic loops with decreasing temperature -- if the heating
dominates the decay.  This approach has been tested on a sample of
solar flares observed with Yohkoh/SXT (Reale et al. 1997) and
extensively applied to flares observed on stars of various spectral
type (Reale \& Micela 1998, Favata et al. 2000, 2001, Maggio et al.
2000, G{\" u}del et al. 2001).

The empirical methods are of easy application and appropriate to infer
the size of the flaring loops and some information on the flare
heating, provided that the light curve and the temperature and emission
measure diagnostics are available with enough photon statistics and
time resolution and coverage to derive a decay trend. 

An XMM-Newton observation of the nearest star Proxima Centauri, of
spectral type dMe, includes a very well-observed flare, already
presented in G\"udel et al. (2002, hereafter Paper I) and further
analyzed in G\"udel et al. (2003, hereafter Paper II).  Other flares
have been observed on Proxima Centauri and studied in detail (Haisch et
al.  1983, Reale et al.  1988, Poletto et al. 1988, Byrne \& McKay
1989). However, the large effective area and the high time coverage of
XMM-Newton has allowed to collect data at an unprecedented level of
detail which motivate a deeper analysis aimed at a higher level of
diagnostics.

Time-dependent hydrodynamic loop models have been shown to provide a
good description of the evolution of the flaring plasma (e.g. Peres et
al. 1987, Reale \& Peres 1995, Hori et al. 1997).  In particular, in
the hypothesis of compact flares, it is customary to assume that plasma
moves and transports energy only along the magnetic field lines, and to
consider one-dimensional models (e.g.  Nagai 1980, Peres et al. 1982,
Doschek et al.  1983, Nagai \& Emslie 1984, Fisher et al.  1985,
MacNeice 1986, Gan et al. 1991).

The light curve of the flare observed with XMM-Newton shows a very
peaked maximum and a globally slow decay, with changing e-folding time
and even a well-defined smoother secondary peak.  In this work, the
flare is modelled in detail throughout the late phases, well after the
second peak.  The high quality of the data and their detailed
comparison to the model results allow us not only to constrain the
length of the main flaring loop, but also to diagnose the involvement
of other structures in the flare, and to constrain the heating
functions (intensity, temporal and spatial distribution) of all the
flaring structures.

In Sect.~\ref{sec:obs} we describe the observation and the constraints
on the modeling, in Sect.~\ref{sec:model} we describe our modeling
approach in detail; in Sect.~\ref{sec:res}, the simulations performed
and the results obtained are presented, in Sect.~\ref{sec:disc} the
results are discussed and in Sect.~\ref{sec:concl} conclusions are
drawn.

\section{The observation}
\label{sec:obs}

\begin{figure}
\centerline{\psfig{figure=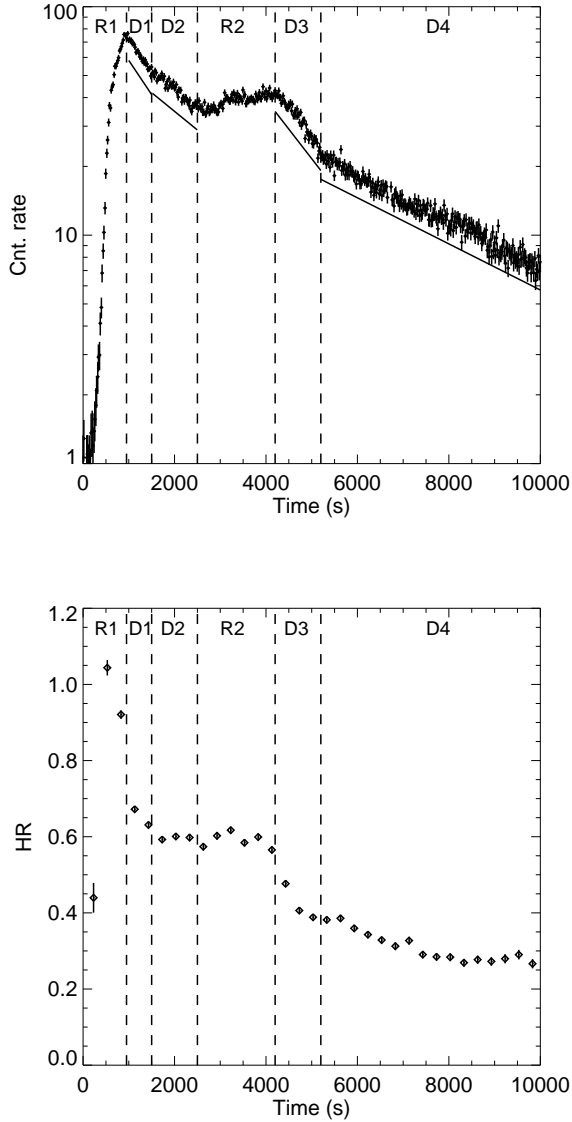,width=8cm}}
\caption[]{{\it Upper panel:} 
Light curve (10 ks) of the flare on Proxima Centauri on 12
August 2001 as detected with the XMM-Newton EPIC-PN detector in the
0.15-10 keV band.  The flare can be segmented into six phases, two
rising (R1, R2) and four decay ones (D1-D4), bounded by the
vertical dashed lines. 
The solid lines mark the decay trends. Time t=0
corresponds to 17:00 UT of 12 August 2001. {\it Lower panel:} hardness ratio
(ratio of 1-4.5 keV to 0.4-1 keV count rates) in the same time interval as
the light curve. Time resolution is 300 s.
\label{fig:datlc}}
\end{figure}

The flare has been detected during the observation of Proxima Centauri
made by XMM-Newton (Jansen et al. 2001) on 2001 August 12, with a total
exposure time of 65 ks.  Fig.~\ref{fig:datlc} shows the flare light
curve in the  0.15 - 10~keV band, collected, in small window mode, with
the PN detector (Str\"uder et al. 2001) of the European Photon Imaging
Camera (EPIC, Turner et al. 2001); MOS detectors are affected by pileup
problems. High resolution X-ray spectra between 0.35 and 2.5 keV, taken
simultaneously with the Reflection Grating Spectrometers (den Herder et
al. 2001), are also available, and in particular OVII 22~\AA~and Ne IX
13.5~\AA~He-like lines have been detected during the flare and analyzed
in detail, yielding density estimates (Paper I). All data were analyzed
using the XMM-Newton Science Analysis System (version 5.3, for RGS data
version 5.4.1). The flare covers most of the final 20 ks of the
observation with a maximum luminosity $L_{X,0.15-10} \approx 3.9 \times
10^{28}$ erg/s.  A large optical burst was captured with the Optical
Monitor (Mason et al. 2001) in the rise phase of the X-ray flare, and
may be a tracer of the production of non-thermal particles in the
corona (Paper~II).

\begin{figure}
\centerline{\psfig{figure=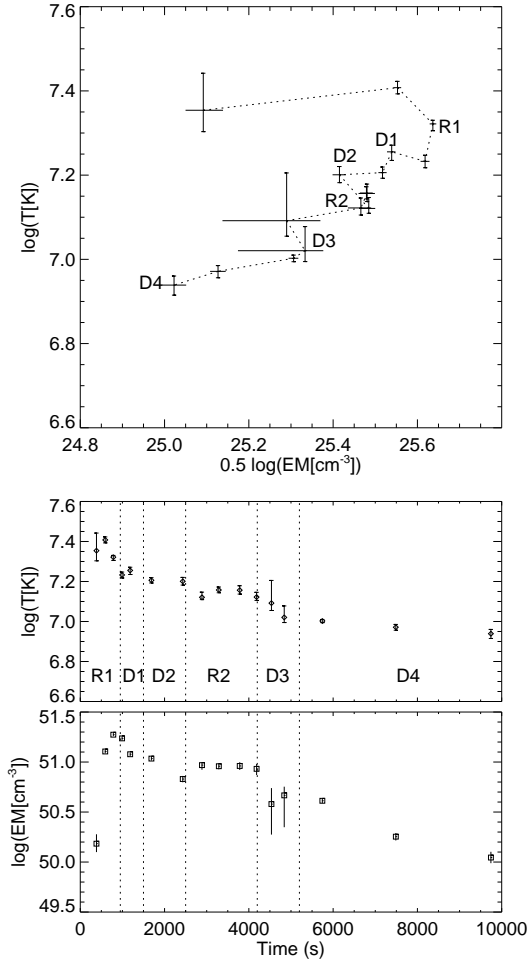,width=7cm}}
\caption[]{Temperature and emission measure diagrams during the flare: the
upper panel shows the density-temperature (n-T) diagram of the dominating
component of the 2-T fitting, where
EM$^{1/2}$ has been used as proxy for the density. The dashed line
marks the evolution of the values and the end point of each phase is
labelled.  The lower panels show the evolution of the temperature and
emission measure separately. Dashed lines as in Fig.~\ref{fig:datlc}.
\label{fig:datnt}}
\end{figure}

Fig.~\ref{fig:datlc} shows the light curve of the first 10 ks of the
flare, and the hardness ratio (ratio of 1-4.5 keV to 0.4-1 keV count rates)
in the same time interval (see also Paper II).
The count rate in the figure is the raw one extracted from the
selected image region which includes 90\% of the total counts and should
be further multiplied by the dead-time correction factor of 1.41.  The
bin size is 20 s.  The corresponding PN spectra, collected in 16 time
intervals as shown in Fig.~\ref{fig:datnt} have been fitted with 2-T
MEKAL models (Mewe et al. 1995) in XSPEC (Arnaud 1996).
Fig.~\ref{fig:datnt} shows the values obtained for the dominant hotter
component throughout the flare.

The count rate reaches values as high as $\sim 122$ cts/s (after
corrections).  We can identify different phases of the light curve,
which will be relevant for the modeling.  During the initial rising
phase (hereafter R1), the emission increases steeply by about two
orders of magnitudes reaching a maximum in $\sim 1$ ks.  The following
decay is initially relatively rapid (D1, with a duration of $\approx
0.5$ ks) and then becomes slower (D2, $\approx 1$ ks). The time of the
switch from D1 to D2 coincides with the time when the hardness ratio
stalls and changes to a constant level (Fig.~\ref{fig:datlc}). After
$\approx 2.5$ ks since the beginning of the flare, the light curve
rises again (R2), slowly, for $\approx 1.7$ ks, reaching a second
smoother peak at $\sim 40$ cts/s. The following decay is similar to the
one after the first peak: fast first (D3, $\approx 1$ ks) and then
slower (D4, $\approx 5$ ks).  The light curve in phases D1, D2, D3 and
D4 can be reasonably approximated with exponentials, with e-folding
times $\tau_{D1} = 1.45 \pm 0.08$ ks, $\tau_{D2} = 2.66 \pm 0.12$ ks
$\tau_{D3} = 1.62 \pm 0.06$ ks and $\tau_{D4} = 4.35 \pm 0.06$ ks,
respectively, shown in Fig.~\ref{fig:datlc}.

Fig.~\ref{fig:datnt} shows that the maximum fit temperature of the hot
component, $\log (T_{obs}) \approx 7.4$, is reached in phase R1, somewhat
earlier than the maximum of its emission measure $\log (EM) \approx
51.3$. 

In the hypothesis that the
bulk of the flare, the first peak, occurs inside a single flaring loop,
the e-folding time of the light curve and the slope of the n-T path in
the initial decay can be used to estimate the loop half-length,
according to Eq.~(\ref{eq:lreale}), 
calibrated for the XMM-Newton EPIC/PN spectral response, already
applied to a few events (G\"{u}del et al. 2001, Stelzer et al.  2002,
Briggs \& Pye 2003), with:

\begin{equation}
F(\zeta) = {c_a} \exp (-\zeta/\zeta_a) + q_a 
\label{eq:fzeta}
\end{equation}

\noindent
where
\[
c_a = 11.6 \pm 0.5 ~~~~~  \zeta_a = 0.56 \pm 0.06 ~~~~~ q_a = 1.2 \pm
0.1 
\]
\[
~~~~~~~~~~~~~~~~~~~~~~~~~~~~~0.4 < \zeta \leq 1.9
\]

An expression for the
loop maximum temperature $T_7=T_{max}/(10^7$ K) can be derived from
fitting hydrostatic model loops with isothermal models:

\begin{equation}
T_{max} = 0.184 ~ T_{obs}^{1.130}
\label{eq:t7}
\end{equation}
We obtain $\log (T_{max}) \approx 7.6$.

From Fig.~\ref{fig:datnt}, it can be noted that the initial decay D1 is
very short and only two points are defined in the n-T diagram, too few
to obtain a well-defined decay trend. If one assumes the initial decay
D1 is entirely due to plasma cooling, with negligible heating, the
empirical expression of loop length (Eqs.~(\ref{eq:lreale}),
(\ref{eq:fzeta}) and (\ref{eq:t7})) yields an upper limit to the loop half
length $L_{up} \approx 1.6 \times 10^{10}$ cm.

The emission measure shows a second peak during phase R2, as the light
curve does, while the temperature is practically flat. Phase D3 includes
just two points with large error bars. Phase D4 is better
defined:  the temperature and EM of the hot component both decay
monotonically, the temperature more slowly.

\section{The modeling}
\label{sec:model}

\subsection{The set up}
\label{sec:setup}

The general approach to model this flare is an evolved version of the
modeling of another flare observed on Proxima Centauri in 1980 with the
{\it Einstein} Imaging Proportional Counter (Reale et al.  1988). At
variance with the previous modeling effort, here the modeling will
include later phases of the flare and more than one loop component,
allowing us to diagnose contributions of flaring structures other than
the main loop and the heating function at late times.

The model assumptions are those typical of a solar coronal flare loop
modeling (e.g. Reale 2002): the flare in each loop is triggered by a
strong heat pulse; the loop is initially at equilibrium (Serio
et al.  1981) at the temperature ($\sim~4$~MK) of an active region
loop, not far from the peak of the EM distribution in quiescent
conditions ($\approx~3$~MK in Paper~II).  The flaring plasma is
described as a fluid confined in a closed semicircular loop with fixed
geometry and constant cross-section, perpendicular to the stellar
surface and unchanged during the flare.  The plasma moves and
transports energy only along the magnetic field lines running parallel
to the loop, and can therefore be described with a single curvilinear
coordinate. The plasma evolution is then described by the
time-dependent hydrodynamic equations of mass, momentum and energy
conservation as done in many previous works (see references in
Section~\ref{sec:intro}), including, as significant physical effects,
the gravity, the compressional viscosity, the radiative losses from
optically thin plasma, and the thermal conduction. The stellar gravity
and radius have been assumed $g_* = 10 g_{\sun}$ and $R_*
= 0.15 R_{\sun}$, respectively (Pettersen 1980, S\'egrensan et al.
2003).  There are two external energy inputs: a low, constant and
uniform one, which keeps the loop initially at equilibrium; a high and
highly transient one, $Q(s,t)$, which triggers the flare, and is
assumed to be a separable function of space $g(s)$ and time $f(t)$
(e.g. Peres et al.  1987):

\begin{equation}
Q(s,t) = H_0 ~ f(t) ~ g(s)
\end{equation}
where $H_0$ is the peak value of the heating rate,
$s$ is the coordinate along the loop, $t$ is the time. We
consider Gaussian spatial distributions, centered on $s_0$ and with
width $\sigma$:

\begin{equation}
g(s)=\exp \left[ -\frac{(s-s_0)^2}{2\sigma ^2} \right]
\end{equation}
There is no reliable way to determine a priori the intensity, the
spatial distribution, the duration and the time dependence of the
heating function. We therefore proceed by educated guesses and refine
the choices with the feedback coming from the comparison of the data to
the model results.  We will consider, in particular, three alternative
distributions $g(s)$: two thin Gaussians centered at the footpoints, a
single wide Gaussian centered at the apex, and a uniform heating
($\sigma \gg L$).  As for the time dependence, we will consider a heat
pulse, described as $f(t) = 1$ for $0 < t \leq \delta t_H$, and $f(t) =
0$ at any other time.  The heating decay is assumed exponential:

\begin{equation}
f(t) = \exp [-(t-t_d)/\tau_H]
\label{eq:hdecay}
\end{equation}

The time-dependent hydrodynamic equations have been solved using the
revised version of the Palermo-Harvard numerical code with adaptive
regridding (Betta et al. 1997, Betta et al. 2001). Symmetry with
respect to the apex has been assumed, and a half-loop modelled.

A flaring loop model is set up by selecting the loop length and the
heating function.  The details and conditions of the loop before heat
ignition are not critical for the simulation results, provided that the
pressure is high enough to have a significant amount of mass in the
chromosphere for evaporation (see Sect.~\ref{sec:res}).

\subsection{The analysis of the results}

The numerical solutions of the 1-D hydrodynamic plasma equations are in
the form of plasma density, temperature and velocity distributions
along the loop at progressing times. For comparison with observational
data, from each density and temperature distribution, the plasma X-ray
spectrum at the focal plane of the EPIC-PN detector is synthesized as
done in several previous works (e.g. Reale et al. 1988, Reale et al.
1997, Reale \& Micela 1998).  We consider the MEKAL spectral code (Mewe
et al. 1995) with a metallicity Z=0.5, as on average found in the
spectral fits (Paper~II). The MEKAL spectra are folded with the EPIC-PN
response function used in G\"udel et al.  (2001). The final results are weakly
dependent on the details (or minor changes) of the response function,
since they are mainly based on the analysis of global observables, such
as the light curve in a broad spectral band (0.15 - 10 keV).  The
normalization of the light curve obtained from the loop model to the
observed one provides the loop cross-section area, which is a free
parameter in the model.

The loop model spectra are fitted with single temperature (1-T) model
spectra (the same as those used to synthesize them).  The fitting
provides a best-fit ``average'' temperature $T_{fit}$, and an
analytical normalization factor, which, multiplied by the loop
cross-section area, yields the emission measure.  The 1-T fitting is
performed in the 0.8-10 keV sub-band. This allows us to compare
$T_{fit}$ to the temperature of the hot component of the
multi-temperature fitting of the data, at least for $T_{fit} \ga 10$
MK.

Whenever fitting the observation data requires the combination of two
model loops, we synthesize the total emission at a given time by
summing the two focal plane spectra -- one for each loop, with the
appropriate cross-section area -- at that time. We then analyze the
resulting sequence of spectra, one for each time, as we do for single
loop spectra:  we derive the light curve and fit the spectra with
single temperature models.

\section{The results}
\label{sec:res}

The modeling of this flare will be described following the flare
evolution.  It will first address the flare peak, i.e.  phases R1 and
D1 in Fig.~\ref{fig:datlc}, then the first decay (D2), and finally the
second peak and the late decay (R2, D3 and D4).

\subsection{The flare peak}

We model the initial and most intense phase of the flare with a single
flaring loop; we will call it {\it loop A}.  

\subsubsection{The length of loop A}
\label{sec:llen}

The modeling requires, first of all, that we set the loop length.  As
mentioned in Section~\ref{sec:obs}, the empirical scaling laws applied
to phase D1 provide an upper limit $L_{max} = 1.6 \times 10^{10}$ cm.
In the lack of a well-defined path in the n-T diagram, and therefore of
reliable information about the heating decay, any length shorter than
this may be appropriate.

We will show here results for three loop half-lengths, namely the upper
limit $L = 1.6 \times 10^{10}$ cm, an intermediate value $L = 1.0
\times 10^{10}$ cm and the half-length obtained for the {\it Einstein}
flare ($0.7 \times 10^{10}$ cm, Reale et al.  1988). The initial base
pressures are $p_0 = 3$ dyne cm$^{-2}$ for the first two, and $p_0 =
4.3$ dyne cm$^{-2}$ for the last length value.

\subsubsection{The heat pulse}
\label{sec:heat}

The flare peak is driven by a strong heat pulse.  The time dependence
of the heat pulse is described in Section~\ref{sec:setup}. The data
indicate a very rapid increase of the temperature and therefore an
impulsive heating.  Typically in flares (and in their simulations as
well) the emission measure still increases well after the heating has
been turned off (e.g. Svestka 1976). The temperature is a better tracer
of the heating duration, because the efficient thermal conduction makes
it promptly decrease as the heating decreases. Fig.~\ref{fig:datnt} and
the time evolution of the hardness ratio (Fig.~\ref{fig:datlc}) suggest
a duration of the order of 500 s; the choice of a pulse duration
$\delta t_H = 600$ s is good for all our simulations of this flare
phase.

A hint for the pulse intensity comes from the flare maximum
temperature. It is reached after a few seconds and then remains steady
as long as the heating is constant, because thermal conduction rapidly
balances the heating.  By applying the loop scaling laws (Rosner et al.
1978) with $\log T = 7.6$ (Section~\ref{sec:obs}), we obtain that, if
the heating were distributed uniformly in a loop of half-length
$10^{10}$ cm, its intensity would be of the order of:

\begin{equation}
E_H \approx 10^{-6} T^{3.5} L^{-2} \approx 4 ~ \rm erg ~ cm^{-3} ~ s^{-1}
\end{equation}

For this phase of the flare, we have considered two alternative spatial
distributions of the heat pulse along the loop: i) just above the
footpoints ($s_0 = 0.1 L$ and $\sigma = 0.03 L$); ii) centered at the
apex ($s_0 = L$ and $\sigma = 0.3 L$).  The width of the heating
distribution influences the simulation results little.  

\subsubsection{Modeling phases R1 and D1}

\begin{figure*}
\centerline{\psfig{figure=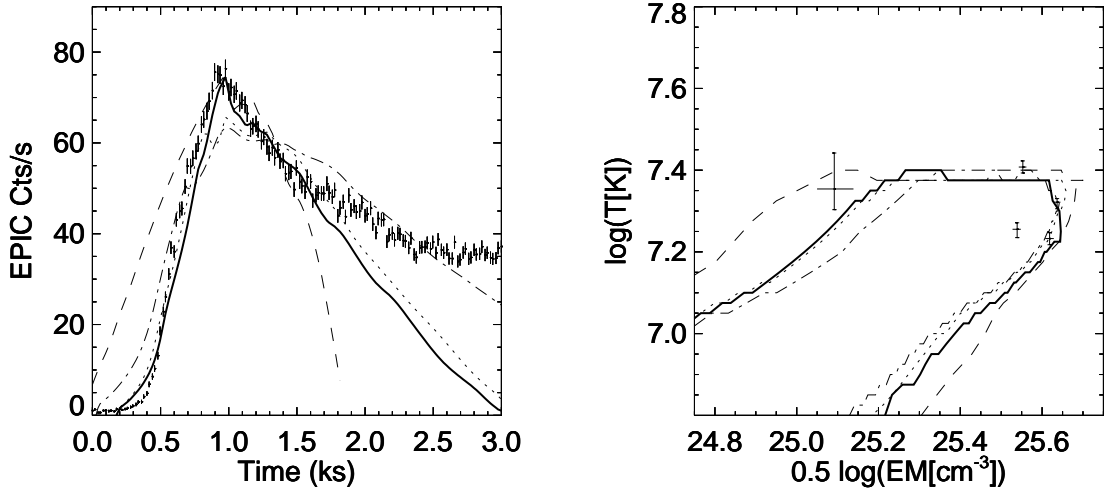,width=16cm}}
\caption[]{{\it Left:} Fitting the observed light curve ({\it data
points}) of the phases R1 and D1 with hydrodynamic simulations of a
single flaring loop.  The figure shows the light curves obtained from a
loop with half-length $10^{10}$ cm heated at the footpoints ({\it solid
line}) and at the apex ({\it dotted line}), from a loop with
half-length $0.7 \times 10^{10}$ cm heated at the footpoints ({\it
dashed line}), and from a loop with
half-length $1.6 \times 10^{10}$ cm heated at the apex ({\it
dashed-dotted line}). {\it Right:} corresponding paths in the EM$^{1/2}$-T diagram, obtained from 
fitting the model and the observed spectra with isothermal models
(only the first five data
points are shown).
\label{fig:fitr1d1}}
\end{figure*}

We will not report on the whole exploration of the space of the model
parameters that we performed, but only on some cases providing
representative results.  We will discuss results for the three loop
lengths listed in Section~\ref{sec:llen}.  For the intermediate loop
length ($L=10^{10}$ cm), we show results for two cases, i.e. a heat pulse
concentrated at the loop footpoints with a maximum intensity of $H_0 =
60$ erg cm$^{-3}$ s$^{-1}$, and a heating deposited at the loop apex
with a maximum intensity $H_0 =  12$ erg cm$^{-3}$ s$^{-1}$. For the
smallest length, we show results for a heat pulse concentrated at the
loop footpoints with a maximum intensity of $H_0 = 85$ erg cm$^{-3}$
s$^{-1}$. For the longest loop, we show results for a heat pulse at the
loop apex with a maximum intensity of $H_0 = 10$ erg cm$^{-3}$
s$^{-1}$.

The evolution of the flaring plasma confined in a loop is well-known
from extensive previous modeling studies (e.g. Peres et al. 1982). The
global characteristics of the evolution do not depend on the details
of the heating (see also Section~\ref{sec:evol}): the heat pulse
makes the temperature increase up to several tens MK along the whole
loop in a few seconds, due to the high plasma thermal conduction; the
dense chromosphere at the loop footpoints is heated violently, and
expands upwards with a strong evaporation front. The upcoming plasma
fills up the loop, very dynamically first and then more gradually,
approaching a new hydrostatic equilibrium at a much higher pressure.
The loop X-ray emission increases mostly following the increase of
emission measure.  As the heating stops (or decreases), the temperature
promptly begins to decrease everywhere in the loop. As mentioned in
Sect.~\ref{sec:heat}, the emission measure peaks later and then
decreases too, with the timescale of the plasma cooling. The different
modeling choices lead to different time scales, values of density and
temperature, and to different details of the evolution, which, in turn,
determine differences in the X-ray emission and its evolution.

Fig.~\ref{fig:fitr1d1} shows the light curves obtained from computing
3000~s of plasma evolution for the four representative
models described above (two
for the intermediate loop, one for the short loop and one for the long
loop). The model results
are sampled with a minimum sampling time of 20 s. The model light curves
are matched to the data by synchronizing the light curve maxima. The
loop cross-section areas obtained from the normalization of the light
curves are 3.6, 4.2, 1.1 and 4.0 $\times 10^{18}$ cm$^2$, corresponding to
aspect ratios $R/L = 0.11, 0.12, 0.08$ and 0.07, respectively, where $R$ is the
radius of the loop cross-section, assumed circular.

The light curves all rise steeply during chromospheric evaporation, and
the steepness decreases when the evaporation becomes more gradual. The
maximum occurs about 400 s after the heat pulse has stopped, and the decay
follows the decrease of the emission measure due to plasma cooling.

The rising phase obtained from modeling the shorter loop\footnote{A
similar light curve is obtained with heating at the apex.} is too slow
to fit reasonably well the observed one. The emission evolution
obtained with the long loop is too gradual around the flare maximum,
due to the longer time scales implied, and is not able to describe the
sharp flare peak\footnote{A similar light curve is obtained with
heating at the footpoints.}.

The light curves obtained with the intermediate loop fit better
both the rising and the peak phase.  The heating at the footpoints fits the
maximum better than the heating at the apex.  The rise is slower with
the shorter loop because the initial evaporation front takes less time
to fill the loop, and, since then, the density -- and the X-ray
emission -- increases more gradually.

The T vs EM$^{1/2}$ plot of Fig.~\ref{fig:fitr1d1} shows the paths obtained
from fitting the spectra of the four flare models, at various times,
to isothermal model spectra, in comparison with fittings of the data
(first five points).  All models match reasonably well the first four data
points, all included in phase R1 and D1. They depart from the fifth
data point, which belongs to a later phase, as the light curves do
after time $t\sim 1500$ s, and exactly where the hardness ratio stops
decaying and gets constant (Fig.~\ref{fig:datlc}).

The results shown so far suggest us that the model which best fits phases
R1 and D1 of the flare is the one with the loop of intermediate length
($L=1.0 \times 10^{10}$ cm) and the heating at the footpoints. We will
consider this as the starting point for fitting the following phases.

\subsection{The decay phase D2}
\label{sec:d2}

In phase D2, the decay of the light curve slows down significantly, as
shown in Fig.~\ref{fig:datlc}. This trend cannot be explained with the
cooling of a longer loop (Eq.~(\ref{eq:tauserio})), because the decay
is initially faster. Nor can it be explained with a heating gradually
decaying from the peak value: there would be a single slower decay
trend. 

This change of slope may be explained in two alternative ways:  i) in
loop A, a low residual heating, much lower than the initial impulsive
heating and with a different evolution, remains active; ii) another
loop is beginning to flare and its rising emission overlaps the
emission of loop A, slowing down the overall decay, and determining
also the second flare peak (R2$+$D2).

The latter hypothesis will be discussed in the next paragraph,
together with later flare phases. We now explore the 
hypothesis of the residual heating: it must be significantly less
intense than the strong initial pulse, so as to become important only later
in the decay, and must decrease gradually so to drive the late decay.
We neglect the effect of such residual heating as long as the
initial pulse is on.

\begin{figure}
\centerline{\psfig{figure=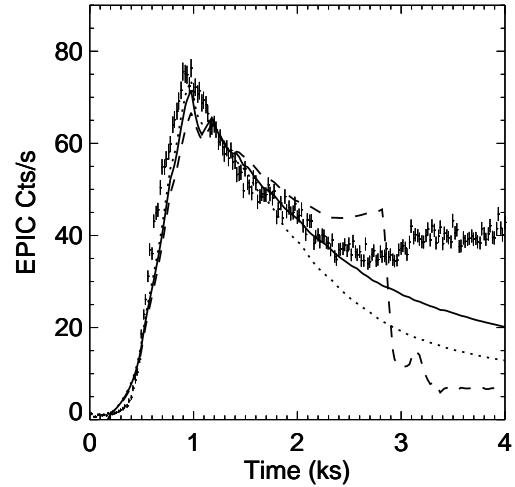,width=8cm}}
\caption[]{Fitting the observed light curve ({\it data points}) of the
flare phase D2 with a decaying heating switched on at the end of the
heating pulse at the footpoints: uniform in corona and initial
intensity 1.2 erg cm$^{-3}$ s$^{-1}$ ({\it solid line}), the same with
initial intensity 0.75 erg cm$^{-3}$ s$^{-1}$ ({\it dotted line}), and
with a decaying heating deposited at the footpoints (15 erg cm$^{-3}$
s$^{-1}$, {\it dashed line}). Data points as in
Fig.~\ref{fig:fitr1d1}.
\label{fig:fitd2}}
\end{figure}

Fig.~\ref{fig:fitd2} shows the fitting obtained with three different
decaying heating functions, each switched on at the end of the heating
pulse deposited at the footpoints, i.e.  $t_{d} = 600$ s and $\tau_{H}
= 4500$ s in Eq.~(\ref{eq:hdecay}):  a) uniform in corona, and with
intensity $H_0 = 1.2$ erg cm$^{-3}$ s$^{-1}$; b) the same, with lower
intensity $H_{0} = 0.75$ erg cm$^{-3}$ s$^{-1}$; c) at the footpoints,
with the same spatial parameters as the impulsive heating and with
intensity $H_{0} = 15$ erg cm$^{-3}$ s$^{-1}$.  Integrating over the
loop length, the total rates of the heating a) and c) are $\approx 1/4$ of the
total rate of the impulsive heating, heating b) $\approx 1/6$.  The
loop cross-section area obtained to best-fit the light curve down to
phase D2 slightly changes to 3.3, 3.5, and $3.2 \times 10^{18}$ cm$^2$,
respectively.

It is immediately apparent that the residual heating deposited at the
footpoints cannot describe the light curve in phase D2. At time
$t \approx 2500$ s a thermal instability occurs, and the light curve
first increases and then suddenly drops: indeed any decaying heating
deposited at the footpoints has been found to be unable to describe this
decay phase, because a thermal instability invariably occurs.

Phase D2 appears instead to be described more adequately with the residual
heating deposited uniformly in corona and as low as 1/4 of the impulsive
heating rate at t~=~600~s. An even lower heating rate makes the
emission decrease too fast. An equal amount of heating more localized
anywhere in the coronal part of the loop  (e.g. at the apex) does not
bring significantly different results, and we will henceforth refer
to a residual heating generically deposited in the corona.

\subsection{The second peak and late decay: phase R2, D3 and D4}

After time $t \approx 2500$ s, the light curve rises again to form the
second lower maximum. The question, of course, is how this maximum is
produced. Occam's razor argument would suggest us simply that a second heating
pulse, weaker than the first one, occurs in loop A around that time.
However, we find that a second heat pulse as intense as to produce
the necessary emission increase in the same loop invariably leads to a
temporary temperature increase; such temperature increase is not
observed (Fig.~\ref{fig:datnt}). On the contrary, the temperature stays
constant in phase D2 and slightly decreases in phase R2. A similar
trend is present also in the time evolution of the hardness ratio
(Fig.~\ref{fig:datlc}).

The only way to produce the second emission peak with no significant
temperature change is that a second loop system gets involved in the
flare, and its emission adds to the one of loop~A.  We call this second
loop, {\it loop~B}, and model it separately from loop A. In addition to
its length and its heating function, for modeling loop~B we have to set
a time shift of the heating switch on with respect to loop~A.  It is
not trivial to constrain all these parameters because the evolution of
loop~B must be decoupled from the decay tail of loop~A.

\begin{figure}
\centerline{\psfig{figure=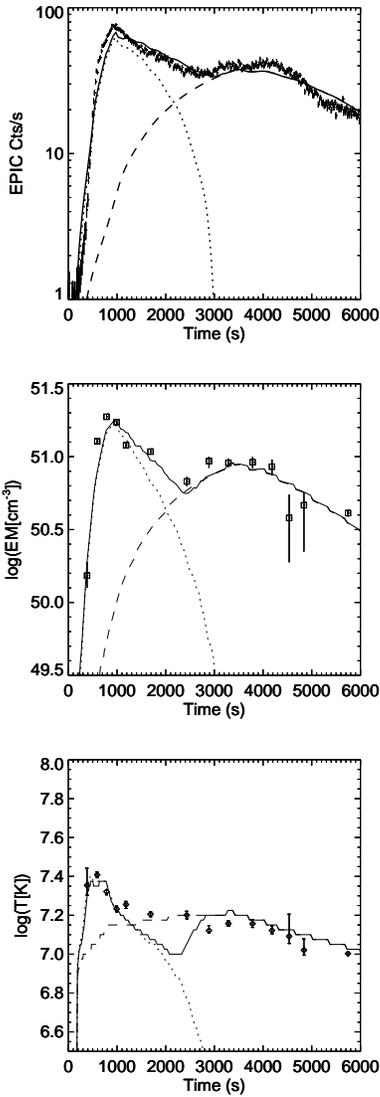,width=10cm}}
\caption[]{Fitting the observed flare light curve ({\it data points})
from phase R1 to D3 with a model consisting of the sum ({\it solid
line} of a footpoint-heated flaring loop with half-length $10^{10}$ cm
(with no decaying heating, {\it dotted line})), and of a second
top-heated flaring loop with half-length $2.5 \times 10^{10}$ cm,
ignited at the same time as loop~A ({\it dashed line}).
The two lower panels show the emission measure and temperature versus time
obtained from 
fitting the model and the observed spectra with isothermal models. 
Data points as in Figs.~\ref{fig:datlc} and \ref{fig:datnt}.
\label{fig:lntp}}
\end{figure}

We start noticing that phases D3 and D4 of the light curve
(Fig.~\ref{fig:datlc}) are similar to phases D1 and D2: the decay is
initially faster and then slows down. The e-folding time in phase D3 is
only slightly longer than the one in the corresponding D1 phase
(see Sect.~\ref{sec:obs} and Fig.~\ref{fig:datlc}). We
take this as an indication that loop B may be similar to loop A, i.e.
same length, and we choose to check this assumption against the
possibility of a much longer loop B, namely L=2.5 $\times 10^{10}$ cm.
Indeed, loop B could be even shorter than loop A, if a residual heating
were present during phase D1. However, this would imply two different regimes
of residual heating, one in phase D1 and another in phase D2, and introduce
another set of free parameters in the modeling, which we prefer not to do,
if unnecessary.

Since the slower evolution expected from a longer loop may naturally
lead to a delayed emission peak, we have explored the possibility that
the flare in this long loop B is triggered at the same time as that in
loop A, with a lower intensity and longer duration. In this hypothesis,
the slower decay D2 may be explained with the superposition of the
continuation of the fast decay D1 with the rise of the flare of loop B
(hypothesis (ii) in Sec.~\ref{sec:d2}).
{\it In this specific scenario, we will drop the decaying heating in loop
A.}

Fig.\ref{fig:lntp} shows the results obtained with a heating pulse
located at the top of loop B with $H_0 = 1$ erg cm$^{-3}$ s$^{-1}$,
constant for 3200 s. The figure shows the light curve of the two flare
components separately, and the light curve obtained by summing
the spectra of loop~A and loop~B (with distinct
cross-section areas) at corresponding
times, compared to the observed light curve.  The light curve of loop B
rises very gradually and peaks at time $t \approx 3500$ s, more than
2000 s later than the flare in loop A.  The loop cross-section of this
second loop that best fits the light curve is $6.2 \times 10^{18}$
cm$^2$, which corresponds to quite a small aspect ratio $R/L \approx
0.06$. Fitting the total spectra of loop~A$+$loop~B
with isothermal models, the global evolution of the emission measure is
well described. The evolution of the temperature instead shows a deep
minimum at time $t \approx 2000$ s. This is not present in the data,
and suggests us to reject this model. A similar temperature dip appears
also if such long loop B is heated at the footpoints.

\begin{figure}
\centerline{\psfig{figure=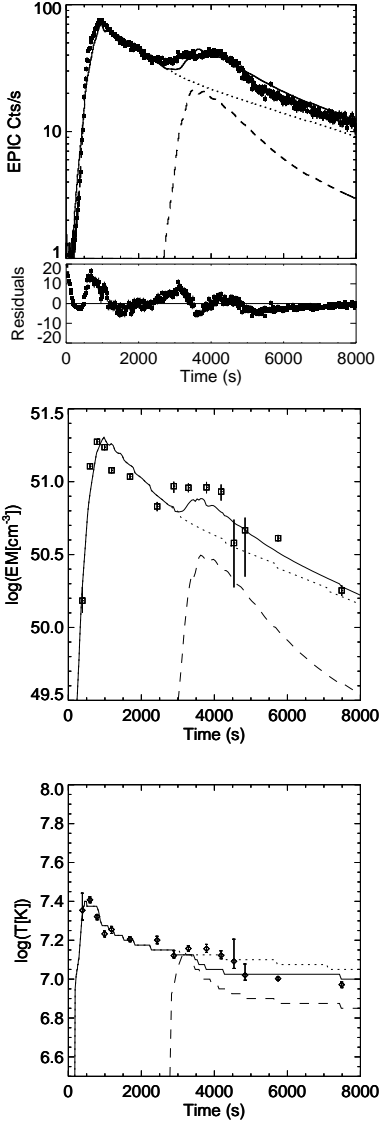,width=10cm}}
\caption[]{Fitting the observed flare light curve ({\it data points})
from phase R1 to D3 with a model consisting of the sum ({\it solid
line}) of two flaring loop systems with the same half-length and
similar heating function: one (loop A, {\it dotted line}) is heated
2600 s earlier and 10 times more intensely than the other (loop B, {\it
dashed line}).  A residual heating sustains the decay of both loops.
Data points and lower panels as in Fig.~\ref{fig:lntp}.
\label{fig:bs6}}
\end{figure}

\begin{figure}
\centerline{\psfig{figure=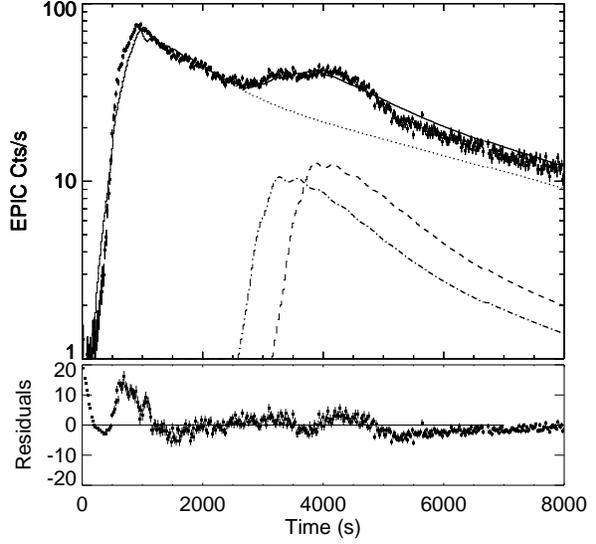,width=8cm}}
\caption[]{Fitting the observed light curve ({\it data points}) of the
whole flare including two other flaring loop components. Residuals (data
counts minus model counts)
are also shown. Data points as in Fig.~\ref{fig:fitr1d1}.
\label{fig:best}}
\end{figure}

Fig.~\ref{fig:bs6} shows results obtained with a loop~B twin of loop~A,
and with a heating duration and spatial distribution of loop B
identical to that of loop A, but triggered 2600 s later, with a
rate ten times lower, $H_0 = 6$ erg cm$^{-3}$ s$^{-1}$. In this
alternative scenario, the decaying heating of loop A is maintained. The
best-fitting cross-section area of loop B is $16.7 \times 10^{18}$
cm$^2$, five times larger than the area of the first flaring loop, or,
equivalently, an arcade of five loops equal to loop A. This combination
describes better the temperature trend, but the light curve shows a
small dip at time $t \sim 3000$ s.  As suggested by the
residuals, the dip can be filled -- and the fit further improved --
simply by adding a third minor flaring component, 
adjusting the loop cross-section areas and the
heating time shifts appropriately (Fig.~\ref{fig:best}). The best
combination that we find is to add two loops~B with area $9.0
\times 10^{18}$ cm$^2$ and $10.8 \times 10^{18}$ cm$^2$ (2.5 and 3
times the area of loop A, still compatible with an arcade of $\sim 5$ loops
equal to loop~A), and heated with time shifts of 2200 s and
2800 s, respectively, since the start of the heating of loop A.

In alternative, we may think to fill the gap of the light curve by
considering a longer-lasting heating pulse in loop B, but we checked
that this choice fails to reproduce adequately the temperature
evolution.

The latest decay D4 can be reasonably well fitted by assuming a
residual heating of loop B (or of two loops B)  with the same
characteristics and e-folding time as the one of loop A, and the
superposition of the decays of loops A and B.

\section{Discussion}
\label{sec:disc}

\subsection{The general scenario}

This work describes the hydrodynamic loop modeling of an X-ray flare
observed on Proxima Centauri with XMM-Newton. The data are very
detailed:  we can distinguish six well-defined phases in the light
curve and a well-defined path in the density-temperature diagram. Our
approach has been to model each phase in detail taking the
time-resolved density/temperature information into account. 

We find that this flare is best-described with the following
combination of components:

\begin{itemize}

\item The initial phase of the flare, including the rise phase, the
main peak and the initial decay, occurs in a single loop, loop A, with
half-length $1.0 \times 10^{10}$ cm.

\item This phase is triggered by a heat pulse deposited at the loop
footpoints over $\sim 10$ min.

\item A (four-fold) weaker residual heating is left in loop A. It is
deposited in the coronal section of the loop and decays slowly with an
e-folding time of $\sim 1$ hour.

\item About half an hour after the first heat pulse has stopped,
other minor heating pulses ignite other loops, probably an adjacent
arcade, and produce a second minor peak in the light curve.

\item The loop arcade is made of $\sim 5$ loops with same length
and cross-section area as loop A.

\item The heating function of the arcade is very similar to that of
the main flare.

\end{itemize}

Table~\ref{tab:loopab} summarizes the parameters of the model
flaring loops. 
\begin{table*}
\caption{Parameters of best-fit loop models}
\label{tab:loopab}
\begin{center}
\begin{tabular}
{l c c}
\hline
Parameters & Loop A & Loop B \\ 
\hline
Geometry&& \\
\hline
Half-length  ($10^{10}$ cm)&    1.0 & 1.0 \\
Cross-section ($10^{18}$ cm$^2$) &3.3 (3.6)$^a$& 17 (20)$^b$ \\
Aspect (R/L) &0.10 (0.11)$^a$&0.22 (0.25)$^b$\\
Morphology&Single loop & Arcade ($\sim 5$ loops)\\
\hline
Heat pulse && \\
\hline
Location & footpoints & footpoints \\
Rate$^c$ ($10^{28}$) erg/s& 27 & 14 \\
Start time (s) & 0 & 2600 (2200 - 2800)$^b$ \\
Duration (s) & 600 & 600  \\
Total energy$^c$ ($10^{32}$ erg) &1.6 & 0.8\\
\hline
Heat decay && \\
\hline
Location & corona & corona \\
Initial rate$^c$ ($10^{28}$) erg/s& 7.2 & 1.8 \\
Start time (s) & 600 & 3200 (2800 - 3400)$^b$ \\
$e$-folding time (s) & 4500 & 4500 \\
Total energy$^c$ ($10^{32}$ erg) &3.2 & 0.8\\
\hline
Plasma parameters && \\
\hline
Max. temperature (MK)&46&20\\
Max. apex density ($10^{11}$ cm$^{-3}$) & 1.1 & 0.2\\
Max. velocity (km/s) & 1400 & 800\\
\hline
\end{tabular}
\end{center}

\noindent
$^a$ - If no residual heating is included in the decay phase.

\noindent
$^b$ - assuming that two arcades of loops B are ignited in a sequence
(600 s one from the other).

\noindent
$^c$ - assuming a loop aspect 0.10 and 0.22, respectively.

\end{table*}

Hydrodynamic modeling allowed us to derive a very detailed scenario
with qualitative and quantitative constraints on the loop morphology
and on the heating function. The modeling may not be unique: we cannot
exclude that other combinations of the parameters may be found with a
significantly higher modeling effort, and a few parameters are not
totally constrained.

On the other hand, we notice that such a complex and detailed event is
reasonably well-described in terms of only two dominant loop components
and a well-defined heating function, valid for both flaring loops. This
result may provide a general pattern for the interpretation of stellar
flares, even those which show light curves more complex than a simple
rise$+$decay.

The complex light curve of this flare may be in part explained by the
larger collecting area of XMM together with its capability of long
uninterrupted observations with respect to previous satellites, and we
may have missed it in other stellar flares just because of insufficient
S/N ratio and time coverage. On the other hand, there are stellar
flares observed with enough time resolution, coverage and statistics,
which are less complex (e.g. van den Oord \& Mewe 1989, Pallavicini et
al. 1990), and also many solar ones (e.g. Sato et al. 2003).

If the data quality were not so high, we would not have been able to
distinguish so many details of the light curve and to address them one
by one, and we would have limited our analysis, for instance, to an
overall application of the empirical scaling law
(Eq.~(\ref{eq:lreale}),~(\ref{eq:fzeta}),~(\ref{eq:t7})) approximating
the decay to a single decay. We would have obtained a decay time
$\tau_{sin} \sim 4.3$ ks, and a slope in the n-T diagram $\zeta_{sin}
\sim 0.5$. With $\log (T_{obs}) \approx 7.4$ (Fig.~\ref{fig:datnt}), we
would have obtained $L_9 \approx 13$, i.e. 30\% larger than the best
value derived with detailed hydrodynamic modeling.  The agreement
between the two approaches is relatively good, also considering that
the slope $\zeta_{sin}$ is close to the lower limit of the
applicability of formula~(\ref{eq:fzeta}), and therefore it is
better to take $L_9$ as an upper limit.

\subsection{The loop morphology}
\label{sec:morph}

The hydrodynamic evolution of the plasma confined inside a single loop
of total length $2 \times 10^{10}$ cm (loop~A) is able to explain the
initial phases of the flare. The later phases, and in particular the
second peak, instead require the ignition of a second loop system. The
modeling tells us that a second longer loop triggered simultaneously to
loop~A can fit the second peak, but not the slow monotonic temperature
decay after the flare maximum.  To fit both, it is necessary to assume
a residual decaying heating in the coronal section of loop~A, and an
arcade of $\sim 5$ loops identical to loop~A triggered $\sim 40$ min
later. We come up therefore with a flare involving a system of almost
identical loops, a single one first, and an arcade later.

\begin{figure*}
\centerline{\psfig{figure=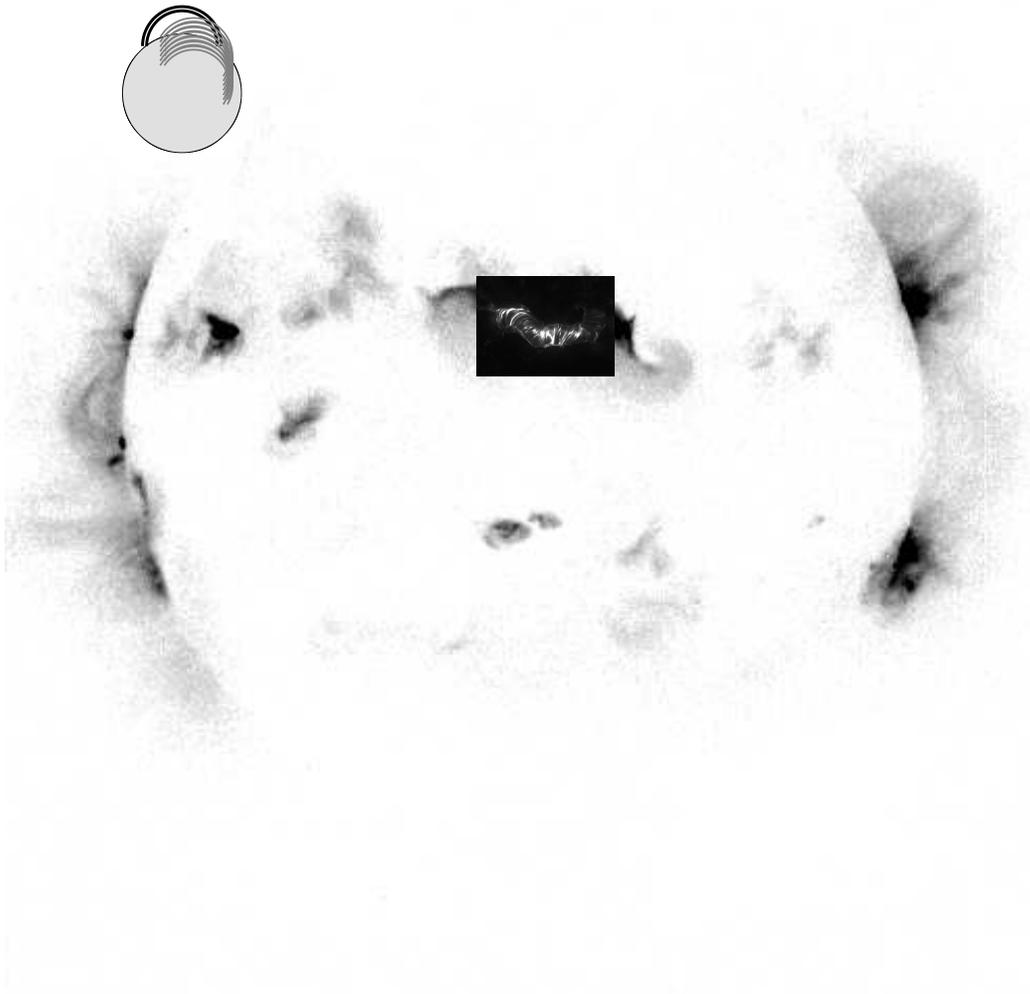,width=14cm}}
\caption[]{Sketch of the possible scenario of the flaring loop system on
Proxima Centauri scaled to the Bastille Day flare on the Sun. The size of
Proxima Centauri and the flare loops are on scale.
\label{fig:photo}}
\end{figure*}

We have also realized that there is at least one solar event which
presents several analogies with this scenario:  the so-called Bastille
Day flare (14 July 2000). This is quite an intense solar flare (GOES class
X5.7) whose light curve in the Al.12 filter passband of the Soft X-ray
Telescope onboard the satellite Yohkoh shows a clear bump after the
main maximum (see Fig.3 in Aschwanden \& Alexander 2001). The same
figure clearly shows also that the bump is associated with the
spectacular ignition of a long arcade, also detected by the TRACE
telescope. All this may suggest a certain similarity of the loop
morphology of this event with that on Proxima Centauri. Also the timing
of the light curve phases is not tremendously different:  the bump of
the solar flare occurs about 600 s after the peak, the second maximum
of the Proxima Centauri flare occurs 3000 s after the first one. The
different delay may be linked to the scale size of the loops:  the
solar arcade loops are a factor $\sim 4$ shorter than the predicted
stellar loops. We sketch a possible scenario of the flaring loop
system on Proxima Centauri scaled to the resolved scenario of the
solar Bastille Day flare in Fig.~\ref{fig:photo}.

The half-length of both loop~A and loop~B is found to be of the same
order as the estimated radius of Proxima Centauri ($L/R_\star \approx
1$). This length is neither very far (1.4 times) from the length
estimated for the flare observed with {\it Einstein} (Reale et al.
1988), nor very large in absolute value (the radius of Proxima Centauri
is indeed a small one), and it is still reasonable as for surface
coverage (as shown in Fig.~\ref{fig:photo}).

\subsection{The flare heating}

The modeling has provided us with detailed information on the heating
deposition, both as for the spatial distribution and for the temporal
evolution.  A heat pulse deposited in the coronal part of loop~A seems
unable to fit the sharp peak of the light curve. A heating deposited at
the loop footpoints is instead more successful.

On the other hand a heating at the footpoints is unable to drive the
observed slow late decay, which instead seems to require a coronal
location. 

The presence of both a steep rise phase and a slow late decay therefore
suggests that {\it both} kinds of heating depositions, one at the
footpoints and the other in corona, must be at work. 

We can also infer the relative weight of the two heating components.
The footpoint heating is more impulsive, i.e. intense and short-lasting
(a few min). The other heating component is less intense ($\sim 1/4$)
and releases its energy over a much longer time scale (one hour). Such
features seem to be traced also by the optical light curve (see Paper
II), which shows a sharp peak and a slower decay starting at $\sim 1/4$
of the maximum optical count rate.

It is interesting to note that: a) the two components contribute to the
flare with comparable amounts of energy ($2 - 3 \times 10^{32}$ erg);
b) a similar combination of these two heating components ($\approx
10^{32}$ erg each) in the loop arcade (loop~B) 
is able to explain the second flare
maximum. Although we do not exclude that refining the heating function
of the arcade may further improve the fitting of the data, we
have shown that using simply the same time parameter values of the
heating function yields a satisfactory description of the flare.

\begin{figure}
\centerline{\psfig{figure=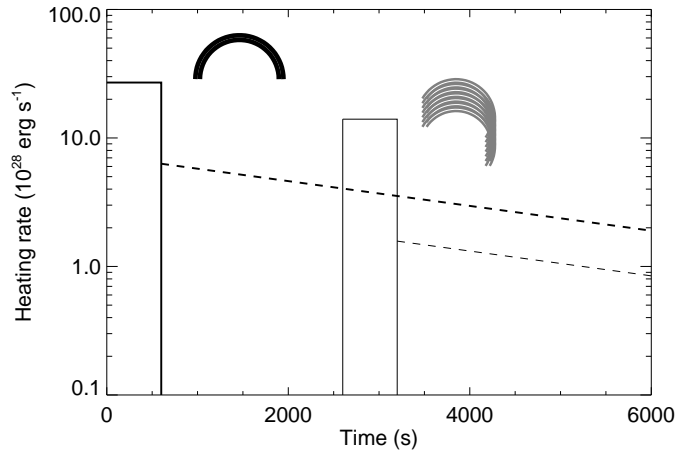,width=7.5cm}}
\caption[]{Heating function of best-fit multi-loop model (shown in
Fig.~\ref{fig:bs6}) of the Prox Cen flare according to our modeling.
The main loop A ({\it black}) is heated first ({\it thick solid line})
with a pulse at the footpoints, followed by a lower and gradual decay
({\it thick dashed line}) deposited in the coronal segment of the
loop.  The arcade of loops B ({\it grey}) is heated later with a
heating function similar to that of loop A ({\it thin solid and dashed
lines}).
\label{fig:heat}}
\end{figure}

The global heating function which drives the flare evolution of
Fig.~\ref{fig:bs6} is shown in Fig.~\ref{fig:heat}.  Given the close
similarity of the heating function of the two flaring structures,
independent although probably adjacent, we may advance the hypothesis
that the two heating components may often be both present in many
flares, and may represent a general characteristics of solar and
stellar flares. The idea that one heating mechanism is probably
insufficient to explain coronal flares is not new (e.g. Peres et al.
1987, Masuda et al. 1994), but our analysis may provide a detailed and
quantitative pattern to be explored in other events.

Note, in particular, that the heating at the footpoints may be driven
by high-energy electron beams precipitating along the loop from a
reconnection site high in the loop, as often mentioned in the
literature (e.g. Masuda et al. 1994), and as also traced by the optical
light curve (Paper II).

\subsection{The energy budget}

On the basis of the model which best describes the X-ray data on this
flare, we can make some considerations about its energy budget.
Considering the loop aspect $R_A/L \approx 0.1$ for loop A and $R_B/L
\approx 0.22$ for loop B, we obtain that the maximum heating rates
injected in the main loop and in the arcade are $\approx 27 \times
10^{28}$ erg/s and $\approx 14 \times 10^{28}$ erg/s, respectively. For
comparison, the maximum luminosity of the flare has been estimated to
be $L_{X, 0.15-10} \approx 3.9 \times 10^{28}$ erg/s (Paper II), i.e.
$\sim 15$\% of the maximum heating rate. The rest of the heating rate
goes into enthalpy, conduction of thermal energy downwards to the
chromosphere, and radiation at lower energies.

The rate of the uniform heating at the beginning of the decays are
$\approx 7.2 \times 10^{28}$ erg/s and $\approx 1.8 \times 10^{28}$
erg/s, respectively. If we integrate in time, we obtain that the total
energy released over the decay phase ($3.2 \times 10^{32}$ erg for loop
A and $0.8 \times 10^{32}$ erg for loop B) is for loop A twice, and for
loop B equal to, the energy released in the heating pulse.

Summing over the whole flare we obtain a total flare thermal energy
input of $6.5 \times 10^{32}$ erg, to be compared to a total energy
radiated in the 0.15-10 keV of $1.5 \times 10^{32}$ erg (Paper I),
i.e.  $\sim 25$\% of the total injected energy. For comparison, the
total energy (of the analyzed part) and the peak X-ray luminosity of
the flare observed on Proxima Centauri with the {\it Einstein}
satellite were $\approx 2 \times 10^{31}$ erg and $\approx 1.2 \times
10^{28}$ erg/s, respectively, i.e.  about 1/3 and 1/4 of the total
energy and peak rate of the heating used in the relevant hydrodynamic
modeling (Reale et al. 1988). That flare appeared to be less energetic
and therefore, consistently, softer and with less efficient thermal
conduction than the flare analyzed here.

For comparison with a solar flare, the total thermal energy involved in
the Bastille day flare has been estimated to be $\sim 0.5 \times
10^{32}$ erg (Aschwanden \& Alexander 2001), about 1/10 of the energy
input of this Prox Cen flare.

\subsection{The plasma evolution and parameters}
\label{sec:evol}

\begin{figure}
\centerline{\psfig{figure=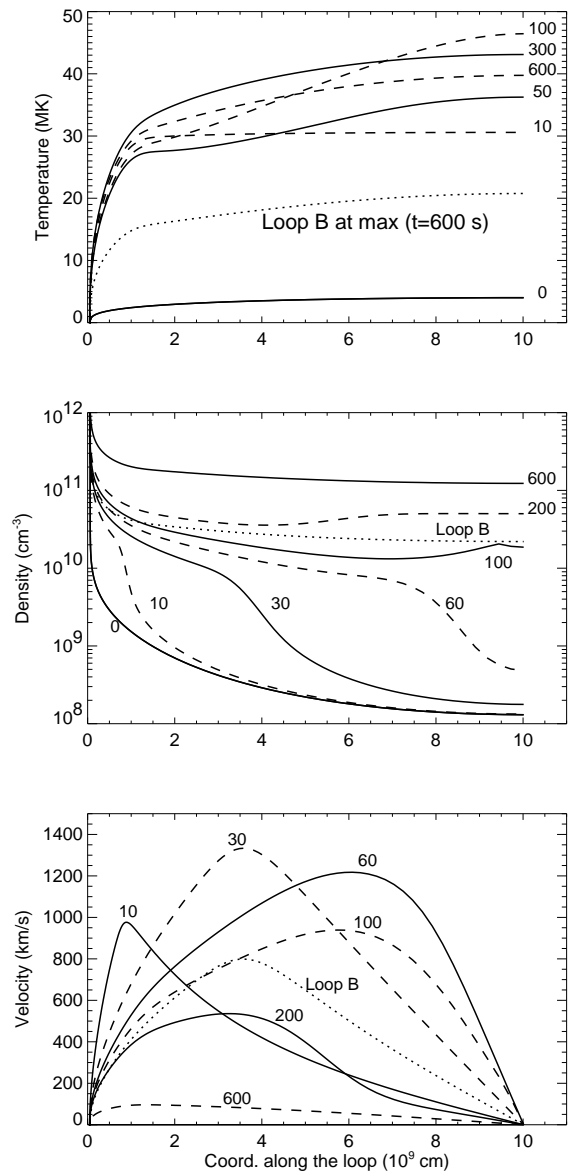,width=8cm}}
\caption[]{Evolution of the temperature, density and velocity along the main
flaring loop. Distributions along half of the loop are shown at 
the labelled times (s). The profiles 
of the second loop system at their maxima are also shown ({\it dotted line}).
\label{fig:profs}}
\end{figure}

The best fitting model provides us with an insight on the evolution of
the plasma involved in the flare and confined in the flaring
structures. Some features are shown in Fig.~\ref{fig:profs} (see also
Table~1). The most dynamic and intense evolution of the flare occurs in
loop A: the plasma temperature rises in few seconds from about 3 MK to
30 MK, and to an absolute maximum close to 50 MK after the first 100 s;
it then settles at about 40 MK during the time the heat pulse is on.
After the first 10 s plasma begins to evaporate significantly upwards
from the chromosphere, with velocities above 1000 km/s and making the
coronal density increase by almost two orders of magnitude to about
$10^{10}$ cm$^{-3}$.  In spite of the high velocity, the relevant
Doppler (blue) shifts would be difficult to be detected even in the 
case of a favourable loop orientation to the line of sight, because
restricted to very hot lines ($> 10$ MK) undetected with the RGS, and a
very short time interval (the initial 2-3 min) of the flare, as
typically occurs in solar flares (e.g. Antonucci et al. 1987). The
blue-shifted line component is therefore lost, because highly diluted in the
typical time intervals of photon integration $\ga 10$ min (Paper I).

After $\sim 1$ min the strong evaporation front reaches the top of the
loop. Since then, the plasma continues to fill up the loop much more
slowly, reaching a density above $10^{11}$ cm$^{-3}$ in the corona.

The loop system B undergoes a similar evolution, but significantly less
dynamic; the maximum temperature is about 20 MK, the coronal density at
its maximum is one order of magnitude less than in loop A, and 
the maximum velocity one half that of loop A.

The modeling highlights the presence of very hot plasma components,
with twice the temperature value than the one obtained from simple data
fitting.

\begin{figure}
\centerline{\psfig{figure=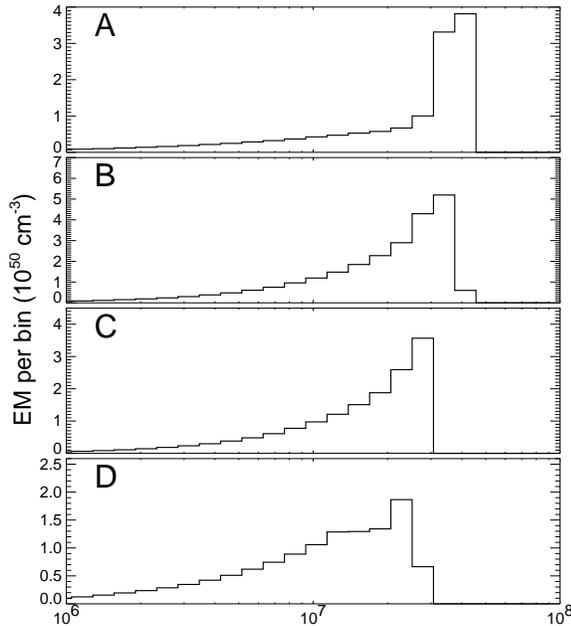,width=8cm}}
\caption[]{Emission measure distributions versus temperature (EM(T))
obtained from hydrodynamic modeling of the flaring loop system, shown
in Fig.~\ref{fig:bs6}. The distributions are averaged over time intervals
corresponding approximately to those of the distributions obtained from data
analysis in Paper~II (intervals A to D).
\label{fig:emt}}
\end{figure}

This is further apparent from the total distribution of the emission
measure versus temperature, EM(T), obtained by summing the
contributions of loop A and arcade B, shown in Fig.~\ref{fig:emt}. The
distributions are averaged over time intervals corresponding to the
ones of the distributions obtained from data analysis, shown in Figs.~8
and 9 of Paper~II (intervals A to D). The EM(T) distributions of
Fig.~\ref{fig:emt} share global similarities with those derived from the
data, in particular to those of Fig.~8 in Paper~II: a dominant hot
component ($\ga 30$ MK) in interval A, a broader and cooler
distribution in interval B, an even cooler distribution with a long
cool tail in interval C, the appearance of a significant cool ($\sim
10^7$ K) component in interval D. The latter cool component relates
to the ignition of the loop arcade B.

The differences between the distributions derived from the hydrodynamic
modeling and those derived from the data are not surprising because the
integral inversion techniques used to derive the distributions in
Paper~II are ill-posed. In spite of this, the comparison with
hydrodynamic modeling clearly provides a key for the interpretation of
the main features of the distributions obtained from the data.

\begin{figure}
\centerline{\psfig{figure=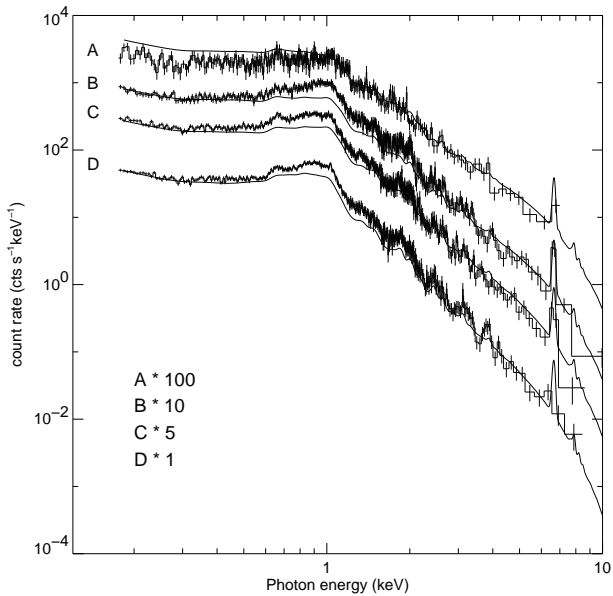,width=8cm}}
\caption[]{Spectra ({\it solid line}) synthesized from results of the best-fit
hydrodynamic loop model (Fig.~\ref{fig:best}) integrated over time intervals
corresponding approximately to those of the spectra shown 
in Paper~II (intervals A to D). Data are overplotted for comparison.
\label{fig:spec}}
\end{figure}

The agreement of the hydrodynamic modeling results to the data is
further confirmed by the focal-plane EPIC-PN spectra synthesized from
the model of Fig.~\ref{fig:best} for intervals A to D, compared to the
observed ones (Fig.~\ref{fig:spec}). The general trends are well
reproduced by the model spectra; discrepancies mainly concern the
intensity of some line groups, mostly related to differences of metal
abundances, which we assume to be Z=0.5 for all elements in the model
spectra. The good agreement in the hard section of the spectra is a
further proof of the presence of significant hot plasma components.

There are no density-sensitive He-like triplets in the RGS band for
very high temperature plasma, and therefore it is difficult to diagnose
the predicted density values of the hottest flaring plasma.  Available
density diagnostics for this flare comes from line ratios of O~VII and
Ne~IX groups obtained with the RGS (Paper I).  The line analysis
provides density values $\approx 4 \times 10^{11}$ cm$^{-3}$ with a
large error bar for O VII, and between $10^{11}$ and $2 \times 10^{12}$
cm$^{-3}$, for Ne IX, in a time interval around the flare peak. In
approximately the same time interval, the loop model yields an average
density of $\approx 2 \times 10^{12}$ cm$^{-3}$, and $\approx 10^{12}$
cm$^{-3}$ at the temperatures of 2 and 4 MK of maximum formation of the
respective ions.  For the Ne~IX, the average density obtained from the
model is compatible to the large interval allowed by the data, although
the Ne IX derived densities are highly uncertain due to severe line
blending (Paper~II). The values obtained from the modeling at 2 MK are
quite higher than those derived from the analysis of the O~VII line.
This may be motivated as follows: the O~VII lines are intensely emitted
both from the flaring plasma and from the remaining quiet corona. What
we detect is therefore the sum of the two contributions, and the
density an average of the 2 MK plasma of the flare and of the whole
Prox Cen corona. If we assume an average density value of the quiet
corona at 2 MK (e.g. $\sim 5 \times 10^9$ cm$^{-3}$, compatible with the
pre-flare density value shown in Paper I), we can infer the
relative weight of the components contributing to form the average
density value and derive an estimate of the plasma volume involved by
the quiet corona component. Just to give an idea of the
volumes involved, this component could be contained in a shell
surrounding the whole Prox Cen star of thickness $\sim 10^{10}$ cm,
i.e. of the order of the stellar radius.

From the properties of the confined plasma, we can also infer some
properties of the magnetic field around the flaring structures.  The
pressure of the flaring plasma confined in loop A reaches values of the
order of 1000 dyne cm$^{-2}$.  In analogy with the derivation in Maggio
et al. (2000), we find that, to keep plasma confined, a minimum
magnetic field of $\sim 150$ G is required; in order to extract a total
energy of $\sim 6.5 \times 10^{32}$ erg in a volume of the order of $5
\times 10^{29}$ cm$^{3}$ and to maintain confinement, the initial
non-potential magnetic field should have been at least $\sim 230$ G.

\section{Final remarks}
\label{sec:concl}

In this work we model a flare observed at a high level of detail
with XMM-Newton on Proxima Centauri. The good time coverage and
resolution and the high count statistics of the data has allowed us to
obtain very detailed diagnostics by means of specific time-dependent
hydrodynamic loop modeling.  The modeling has allowed us, on the one
hand to synthesize in detail a wealth of observables for comparison
with data, and on the other hand to obtain a deep physical insight in
the evolution and distribution of the confined plasma.

On the other hand, the constraints provided by the data allowed us to
discriminate among different model choices, such as different loop
lengths, the presence of more than one flaring loop, the location of
the heat pulse and of the residual heating, their intensity, relative
timing and timescales.  This work indicates that both heating
components are necessary ingredients to explain this flare, and that a
second loop system, probably an arcade, is required to explain the
secondary maximum.

In spite of the high degree of detail and of the many distinct trends
present in the flare, relatively few model components were sufficient to
match the data reasonably: a loop and an arcade of loops, all with the
same length, ignited with some delay by an intense heat pulse at the
footpoints and a gradual residual coronal heating.  
Indeed a solar flare having a similar
evolution, the Bastille Day flare, indicates that the scenario of
involved loops that we find is realistic.  It has been recently shown
that sections of the Bastille Day flare are better described with a
model of several concentric loops than with a single-loop model (Reeves
and Warren 2002). Our results suggest the fundamental ingredients which
govern the X-ray flare evolution are the plasma confinement, a few
dominant loop systems with a fixed length, and a well-defined heating
function. Changes of magnetic topology within each loop system seem to
have a small influence on the X-ray evolution, probably because most of
them are limited to a relatively small fraction of the life of the
flaring arcade (see Fig.~5 in Reeves and Warren 2002).

The loop morphology and the heating function show a well-defined
pattern which may be applied to interpret other stellar flares.  The
heating pattern may be applied, for instance, to stellar flares with
multi-slope decay (Reale 2002 and references therein) and with
secondary maxima (e.g. Poletto et al.  1988, Pallavicini et al. 1990).

It will be certainly interesting also to revisit solar flares showing
similar features under such perspectives and to explore the theoretical
implications concerning, in particular, the flare heating mechanisms.

\bigskip
\bigskip
\acknowledgements{This work was supported in part by Agenzia Spaziale
Italiana and by Ministero della Universit\`a e della Ricerca
Scientifica e Tecnologica. MA and MG acknowledge support from the Swiss
National Science Foundation (fellowship 81EZ-67388 and grant 2000-058827,
respectively). }

{}

\end{document}